\documentclass[12pt]{iopart}
\usepackage[utf8]{inputenc}
\usepackage{amssymb}
\usepackage{amsbsy}
\usepackage{array}
\usepackage{etoolbox}
\AtBeginEnvironment{equation}{\setlength{\arraycolsep}{2pt}}
\usepackage{graphicx}
\usepackage{subfigure}
\usepackage{color}
\usepackage{transparent}
\newcommand{\Eslow}{\hat{\mathcal{E}}_{\rm slow}} 
\newcommand{\Rz}{{\tilde{z}}} 
\newcommand{\RDelta}{\tilde{\Delta}} 
\newcommand{\Rgamma}{\tilde{\gamma}} 
\newcommand{\Rt}{{\tilde{t}}} 
\newcommand{\Rti}{t} 
\newcommand{\Rtau}{\mu\tau} 
\newcommand{\taup}{\tau_{\rm p}} 
\newcommand{\taud}{\tau_{\rm d}} 
\newcommand{\taus}{\tau_{\rm s}} 
\newcommand{\Rtaup}{\Rtau_{\rm p}} 
\newcommand{\Rtaud}{\Rtau_{\rm d}} 
\newcommand{\Rtaus}{\Rtau_{\rm s}} 
\newcommand{\tauR}{\tau_{\rm R}} 
\newcommand{\RtauR}{\mu\tau_{\rm R}} 
\newcommand{\erf}{{\rm erf}}
\newcommand{\PV}{{\rm PV}}
\begin{document}

\title{An efficient quantum memory based on two-level atoms}
\author{Ivan Iakoupov and Anders S. S{\o}rensen}
\address{Quantop -- Danish Quantum Optics center, The Niels Bohr Institute, 
University of Copenhagen, DK-2100 Copenhagen \O, Denmark}
\begin{abstract}
We propose a method to implement a quantum memory for light based on ensembles 
of two-level atoms. Our protocol is based on controlled reversible 
inhomogeneous broadening (CRIB), where an external field first dephases the 
atomic polarization and thereby stores an incoming light pulse into collective 
states of the atomic ensemble, and later a reversal of the applied field leads 
to a rephasing of the atomic polarization and a reemission of the light. As 
opposed to previous proposals for CRIB based quantum memories we propose to 
only apply the broadening for a short period after most of the pulse has 
already been absorbed by the ensemble. We show that with this procedure there 
exist certain modes of the incoming light field which can be stored with an 
efficiency approaching 100\% in the limit of high optical depth and long 
coherence time of the atoms. These results demonstrate that it is possible to 
operate an efficient quantum memory without any optical control fields.
\end{abstract}


\maketitle
\section{Introduction}
Light is an ideal carrier of information both in the classical as well as in 
the quantum regime. To harness the full potential of light for quantum 
information processing it is, however, a major advantage to have access to 
quantum memories capable of storing the information from the light into an 
atomic medium and later releasing the information. Ideally such quantum 
memories should introduce as little disturbance as possible to the information 
encoded in the light field, i.e., the memory should be fully coherent and 
efficient. Furthermore it should be as simple to operate as possible.

\begin{figure}[t]
\begin{center}
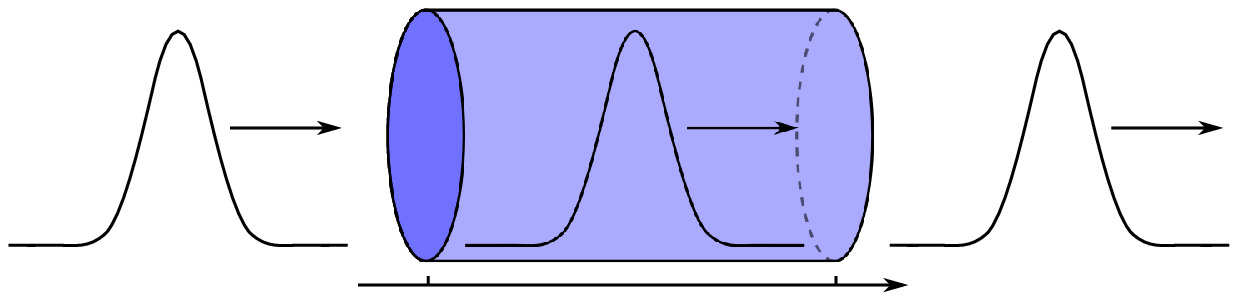
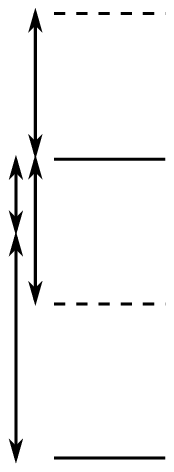
\includegraphics{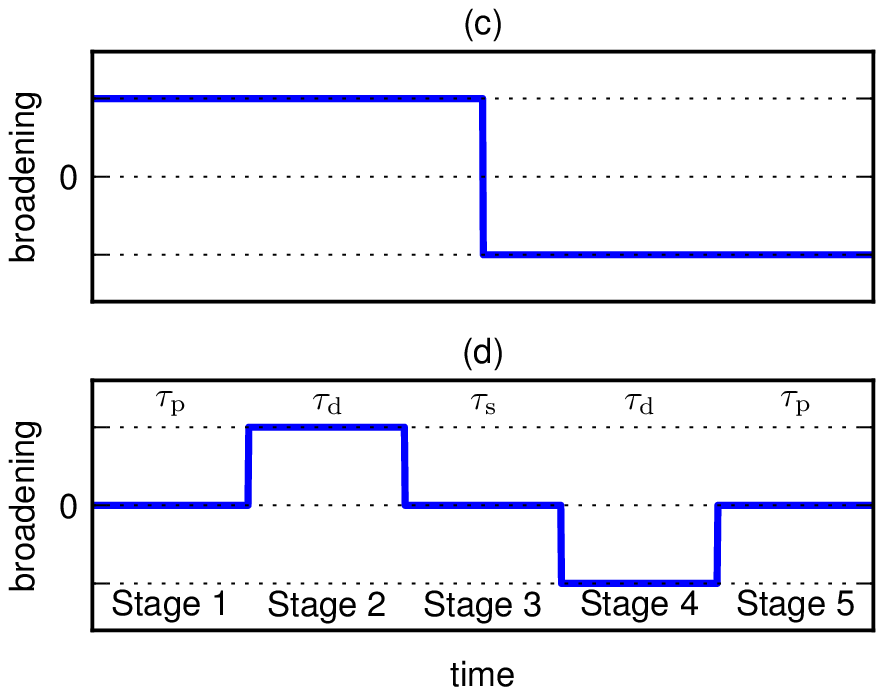}
\end{center}
\caption{Schematic representation of CRIB based memory operation. (a) Setup of 
the quantum memory. The incident light pulse is stored in the atomic ensemble 
of with length $L$ and transverse area $A$ and is later retrieved in the same 
direction. (b) The original atomic transition frequency of the $n$'th atom 
$\omega_{eg,n}=\omega_{\rm L}+\Delta_{0n}$ is only detuned from the electric 
field frequency $\omega_{\rm L}$ by the intrinsic inhomogeneous broadening $\Delta_{0n}$. 
When the controlled broadening is turned on, the original excited state 
$|e\rangle_n$ is further shifted in energy either up 
($|e\rangle_n^{(+)}$) or down ($|e\rangle_n^{(-)}$) so that 
$\omega_{eg,n}=\omega_{\rm L}+\Delta_{0n}\pm \Delta_n$. (c,d) The width 
and direction of the controlled broadening is plotted as a function of time. 
(c) Standard approach where the broadening is applied long before the pulse 
enters the medium. A single reversal of the broadening initiates a rephasing 
which causes a photon echo to be emitted at a later time. (d) The procedure 
proposed here where the broadening is only applied in a short time interval of 
duration $\taud$ after the pulse has already entered the medium during the 
period $\taup$. After the storage time $\taus$ the application of the opposite 
broadening for a duration $\taud$ initiates the retrieval process which last 
for a period $\taup$.}
\label{principle_of_operation_diagram}
\end{figure}

A large number of proposals have been developed for how one can construct 
quantum memories based on atomic ensembles \cite{hammerer}. As opposed to the 
most natural approach of storing one photon in one atom, the idea behind the 
ensembles based approach is to store photons into the collective states of an 
ensemble of atoms. Thereby one avoids the technical challenges associated with 
efficiently coupling single atoms and single photons. As a consequence the 
ensemble based approach considerably simplifies the experimental realization 
of quantum memories. A large class of quantum memory approaches based on 
atomic ensembles uses classical laser fields to control the memory process in 
such a way that the incoming field is mapped into the ground state coherence 
of the atoms. Based on this a number of key experimental advances have been 
achieved (See for instance \cite{stopped-light-2001, Julsgaard2004, 
Eisaman2005, Choi2008, Zhao2008Amillisecond, Reim10, Bloch2009}). Here we 
shall pursue a different approach based on controlled reversible inhomogeneous 
broadening (CRIB) \cite{krausscrib}. The original ({\it transverse}) CRIB is 
developed for impurities embedded in solid state systems. Due to the random 
orientations of the (static) dipole moments of the atoms, an external applied 
field will lead to different shifts in the transition frequency of different 
atoms. Hence an external applied field will dephase the polarization of the 
atoms. This will in essence turn off the reemission of the light absorbed into 
the ensemble which is therefore stored. The central idea in the CRIB approach 
is to later reverse the direction of the external field. If the atomic dipoles 
are fixed, the reversal of the field generates a shift in the opposite 
direction (see figure ~\ref{principle_of_operation_diagram} c). After a 
certain time the polarization will rephase, causing an echo, where the light 
is reemitted. In the original approach an additional ground level was used and 
it was shown that the by driving this additional transition with laser fields 
one can obtain a readout in the backward direction which can in principle 
reach 100\% efficiency (within the theoretical models we will consider here, 
the efficiency is the only parameter characterizing the performance of the 
memory for a single incident mode \cite{hammerer,gorshkov1}, and we will 
therefore focus on this parameter below). From a practical perspective it is 
highly desirable to avoid the use of a third level and the associated lasers 
to drive that transition.

Since the original CRIB proposal \cite{krausscrib} (see also 
\cite{moiseevcrib}) a number of modifications have appeared 
\cite{longcribprl,longcrib,afc} and much experimental progress has been reported 
\cite{earlycribexp, longcribprl, krollexpafc, afcpluslambdaexp, 
hedgesexplongcrib, clausenexpafc, saglamyurekexpafc}. In
particular a different {\it longitudinal} CRIB approach has been developed 
where a gradient of an external field is applied such that the shift depends 
on the position of the atoms \cite{longcribprl}. For longitudinal CRIB it has 
been shown that one can construct an efficient memory based on two level 
systems which in principle can reach an efficiency of 100\% \cite{longcrib} if 
the atoms have a sufficiently long coherence time. For transversal CRIB it was 
also shown that one can construct a quantum memory based on two-level systems, 
but in this case, however, the efficiency was found to be limited to 54\% 
\cite{crib}. Here we develop the theory for an efficient quantum memory based 
on two-level atoms subject to transverse CRIB. We show that by varying the 
broadening in time so that it is not turned on initially but only applied 
during a short time interval as shown in 
figure~\ref{principle_of_operation_diagram} (d), the efficiency of the memory 
can in principle reach 100\% if the optical depth of the ensemble is very high 
and the atoms have a very long coherence time.

\section{Setup and principle of operation}
Our system consists of $N$ two-level atoms that are coupled to a quantised 
electric field. We treat everything in one dimension and thus the electric 
field is only dependent on the spatial coordinate $z$. The atoms are assumed 
to be confined to a length $L$ along the $z$-axis and are distributed with 
uniform density. The transverse extent of the ensemble and cross-section of 
the electric field mode are assumed to be the same with area $A$. See Figure 
\ref{principle_of_operation_diagram} a).

The $n$'th atom has the two internal levels denoted $|g\rangle_n$ and 
$|e\rangle_n$ for ground and excited state respectively. The two levels are 
connected by a transition with a matrix element $\wp$. We assume that the 
atomic transition frequencies are affected by both an intrinsic inhomogeneous 
broadening that we cannot control and a controlled inhomogeneous broadening 
that we can rapidly turn on and off and reverse in a time much shorter than 
the duration of the incoming pulse. These two types of broadening give rise to 
a detuning of the atomic transition from the incoming laser field of frequency 
$\omega_{\rm L}$. The two detunings for the $n$'th atom are denoted by 
$\Delta_{0n}$ for the intrinsic broadening and $\Delta_n$ for the controlled
broadening. Since we reverse the detuning the transition frequency is 
$\omega_{eg,n} =\omega_{\rm L}+\Delta_{0n}\pm\Delta_n$ during the two periods 
where the broadening is applied, see figure 
\ref{principle_of_operation_diagram} b).

We are interested in storing light in two-level systems. Such storage is only 
possible if the lifetime of the excited state is very long. Therefore we shall 
assume that the excited state has a sufficiently long lifetime so that 
spontaneous emission can be neglected. The only source of decoherence of the 
atomic transition which we consider here comes from having an intrinsic 
inhomogeneous broadening of the atomic transition. Note that this assumption 
of negligible spontaneous emission from the excited state does not contradict 
the assumption that light couples to the transition. The quantum light we are 
interested in, couples to collective states which can have a much higher 
emission rates due to collective enhancement effects. The memory can thus have 
a large bandwidth even if the individual atoms have a negligible decay rate. 
Indeed as discussed below, the bandwidth of the memory is characterized by the 
quantity
\begin{equation*}
\mu = \frac{Ng_0^2\wp^2}{\hbar^2}.
\end{equation*}
where $c$ is the speed of light in vacuum and 
$g_0=\sqrt{\hbar\omega_{\rm L}/2c\epsilon_0 A}$. This bandwidth can be related to 
more experimentally accessible variables if we consider a specific model for 
the broadening. Throughout this work we shall assume a Gaussian distribution 
of the intrinsic broadening. In this case the decay of the collective atomic 
polarization due to dephasing is described by a Gaussian in time and we define 
the coherence time $T_2$ as the $1/e$ decay time of the atomic polarization. 
The precise definition and a more detailed discussion can be found in 
\ref{appendix_optical_depth}. The memory bandwidth can then be related to the 
optical depth $d_0$ that will be observed when the controlled broadening is 
turned off, and the coherence time through $\mu=d_0/(T_2\sqrt{\pi})$.

As it was shown in \cite{gorshkovprl,gorshkov2}, $1/\mu$ 
characterizes the fastest time scale for variations in the pulse shape that 
can be allowed if the memory is to store the light with high efficiency. 
For a memory based on three level $\Lambda$-type atoms controlled by a 
classical drive, any pulse shape with a pulse length $\taup\gg 1/\mu$ can be 
stored efficiently in the ensemble. Here we are, however, interested in 
two-level systems. It was also shown in \cite{gorshkov2} that there exist 
certain incoming pulse shapes of duration $\taup\sim 1/\mu$ which can be 
efficiently absorbed and retrieved by having the light interacting only with 
two-level atoms. To control the storage and retrieval process it was proposed 
in \cite{gorshkov2} to rapidly apply a $\pi$-pulse to transfer the population 
from the excited state to an auxiliary state for long term storage after the 
pulse was absorbed in the ensemble. Later another rapid $\pi$ pulse can be 
applied to initiate the retrieval process. It was shown that for the ideal pulse 
shape of the incoming field the efficiency of this memory protocol is 
identical to the highest obtainable efficiency for the memory. Here we shall 
follow a similar approach as the one suggested in \cite{gorshkov2} but without 
the use of a third auxiliary level. Instead we follow a CRIB-like approach 
\cite{krausscrib,crib} and apply a broadening of 
the atomic levels. In our scheme the light pulse is first absorbed in the 
unbroadened ensemble. Afterwards an externally applied inhomogeneous 
broadening rapidly dephases the polarization of the atomic transition. This 
dephasing effectively turns off the reemission of the stored light allowing 
for its long term storage if there is negligible decoherence of the atomic 
state. At a later stage we assume that we can reverse the applied 
inhomogeneous broadening resulting in a rephasing of the atoms which 
effectively turns on the reemission of the pulse. As we shall show below, this 
process allows for an efficient quantum memory if the ensemble has a large 
optical depth.

The memory protocol that we propose can be divided into 5 stages shown in 
figure~\ref{principle_of_operation_diagram}~d). In stage~1 the incoming pulse 
with duration $\taup$ is incident on the atoms and gets absorbed without any 
controlled broadening present. In stage~2 with a (short) duration $\taud$ the 
controlled broadening is applied, which turns off the reemission of the 
excitation. We find that considerably higher efficiencies are obtained if we 
allow the incoming pulse to have a small tail into stage~2 and we thus allow 
this in the detailed simulations below. Stage 3 of duration $\taus$ 
constitutes the main storage period. In this stage we assume that the external 
broadening is turned off. The storage duration $\taus$ is only limited by the 
coherence time $T_2$ of the atomic polarization. In principle this coherence 
time could potentially be prolonged by e.g. using spin echo protocols, but for 
simplicity we shall not consider this possibility here. Stage 4 marks the 
beginning of the read out process. Here the external broadening is reversed 
and applied for a time $\taud$ equal to the duration of the initial broadening 
in stage~2. The read out of the light pulse continues in stage 5 with duration 
$\taup$ where the external broadening is again turned off.

Most approaches to CRIB based memories\cite{crib,longcribprl} do not include 
the stage 3 introduced here but instead use longer stages 2 and 4. The idea 
behind this stage is to increase the flexibility of the readout. If the 
broadening is applied during the whole storage period as in 
figure~\ref{principle_of_operation_diagram}~a) the time it takes from we 
reverse the broadening to the pulse is retrieved is equal to the time from the 
storage until the reversal. Thus a long storage time gives a long response 
time of the memory. On the other hand, as soon as atoms are sufficiently 
dephased in stage~2, the reemission of the light is turned off and any further 
dephasing will not affect the performance of the memory. Hence we might as 
well turn off the broadening after the dephasing time $\taud$ which can be 
much smaller than the storage time $\taus$. Stage 3 cannot be longer than $T_2$ 
if the memory is to be efficient but it can be arbitrarily short. If we want 
to retrieve the excitation stored in the atoms at any point in stage 3 we can 
turn on the reversed broadening (stage 4) and retrieve the excitations after a 
time $\taud$ which is independent of the storage time $\taus$. Additionally, 
small $\taud$ makes it easier to do numerical simulations of the memory as 
discussed below.

\section{Model}\label{Model_section}

In this section we present the model that we shall use to evaluate the 
performance of the memory. Here we shall be rather brief in describing the 
formalism. A more thorough derivation of the equations of motion and the 
principles used is given in \cite{hammerer,crib}. To describe the 
light we consider the electric field operator given in the Schrödinger picture 
by
\begin{equation*}
\hat{\mathcal{E}}(z)
=g_0\Eslow(z)\rme^{\rmi\omega_{\rm L} z/c}
+{\rm H.c.}
\end{equation*}
in terms of its slowly varying components that satisfy
\begin{equation*}
\left[\Eslow(z), \Eslow^\dagger(z')\right]=c\delta(z-z').
\end{equation*}

The atomic transition is subject to two types of broadening, the intrinsic 
broadening and the controlled reversible broadening. The probability 
distribution of the detunings is described by the functions $\mathcal{G}_0$ for 
the intrinsic broadening and $\mathcal{G}$ for the controlled 
broadening. The distributions are assumed to be even functions and are 
normalized according to 
$\int_{-\infty}^\infty \mathcal{G}_0(\Delta_0)\rmd\Delta_0=1$ and 
$\int_{-\infty}^\infty \mathcal{G}(\Delta)\rmd\Delta=1$.

To derive the equations of motion we split the position and detunings of the 
intrinsic and controlled broadening into intervals of size $z_\delta$, 
$\Delta_{0\delta}$ and $\Delta_\delta$ respectively and define the atomic 
polarization operators
\begin{equation*}
\hat{\mathcal{S}}_{klm,ij}=\frac{\mathcal{N}_\mathcal{S}}{N_{klm}}
\sum_{n=1}^{N_{klm}} |i\rangle_n \langle j|_n.
\end{equation*}
Here the sum is over atoms having a position in the interval 
$[k z_\delta, (k+1)z_\delta]$, an intrinsic detuning in the interval 
$[l\Delta_{0\delta}, (l+1)\Delta_{0\delta}]$, and a controlled detuning in 
the interval $[m\Delta_{0\delta}, (m+1)\Delta_{0\delta}]$. The number of atoms 
that have positions and detunings in the mentioned intervals is
$N_{klm} = (N/L)z_\delta
\mathcal{G}_0\left(l\Delta_{0\delta}\right) \Delta_{0\delta}
\mathcal{G}\left(m\Delta_{\delta}\right) \Delta_{\delta}$.
The normalisation constant $\mathcal{N}_\mathcal{S} = (NLg_0\wp)/(\hbar c)$ is 
here chosen such that all the constants of proportionality disappear in the final 
equations of motion below. In the limit 
$z_\delta, \Delta_{0\delta}, \Delta_{\delta}\rightarrow 0$, where 
$\hat{\mathcal{S}}_{klm,ij}$ gets replaced by 
$\hat{\mathcal{S}}_{ij}(z,\Delta_0,\Delta)$. The commutation relation is then
\begin{eqnarray*}
\fl
\left[\hat{\mathcal{S}}_{ij}(z,\Delta_0,\Delta),
\hat{\mathcal{S}}_{i'j'}(z',\Delta_0',\Delta')\right]
=&\frac{\mathcal{N}_\mathcal{S}L}
{N\mathcal{G}_0(\Delta_0)\mathcal{G}(\Delta)}
\delta(z-z')\delta(\Delta_0-\Delta_0')\delta(\Delta-\Delta')
\\
\fl
&\times\left(\delta_{ji'}\mathcal{S}_{ij'}(z,\Delta_0,\Delta)
+\delta_{j'i}\mathcal{S}_{i'j}(z,\Delta_0,\Delta)\right).
\end{eqnarray*}
We shall assume that all atoms are initially prepared in the ground state 
$|g\rangle$ and that only weak fields are incident such that most of the 
population remains in the ground state. When finding commutators we therefore
approximate 
$\hat{\mathcal{S}}_{gg}(z,\RDelta_0,\RDelta)\approx \mathcal{N}_\mathcal{S}$ 
and $\hat{\mathcal{S}}_{ee}(z,\RDelta_0,\RDelta)\approx 0$.
Hence the commutator of $\mathcal{S}_{ge}$ and $\mathcal{S}_{eg}$ is a 
constant, and $\mathcal{S}_{ge}$ can be regarded as proportional to an 
effective harmonic oscillator annihilation operator.

If we introduce the slowly varying operator $\hat{\sigma}(z,\Delta_0,\Delta)
=\rme^{-\rmi\omega_{\rm L} z/c}\hat{\mathcal{S}}_{ge}(z,\Delta_0,\Delta)$ the 
Hamiltonian in the interaction picture and rotating wave approximation can be 
written as
\begin{eqnarray*}
\fl
\hat{H}=\frac{N}{\mathcal{N}_\mathcal{S}L}
\int_{-\infty}^\infty\int_{-\infty}^\infty\int_{0}^{L}
\mathcal{G}_0(\Delta_0)\mathcal{G}(\Delta)\left\{
\hbar\left(\Delta_0+\Delta\right)
\hat{\mathcal{S}}_{ee}(z,\Delta_0,\Delta)\right.\\*
\left.
-g_0 \wp\left[\Eslow(z)\hat{\sigma}^\dagger(z,\Delta_0,\Delta)
+\Eslow^\dagger(z)\hat{\sigma}(z,\Delta_0,\Delta)\right]\right\}
\rmd z \rmd\Delta_0 \rmd\Delta.
\end{eqnarray*}
This Hamiltonian describes the evolution during stage~2 where the controlled 
broadening is turned on. In stages 1, 3, and 5 where there is no controlled 
broadening we can find the Hamiltonian by removing the term with 
$\Delta\mathcal{S}_{ee}(z,\Delta_0,\Delta)$ in the integrand. In stage 4, 
where the detuning is reversed we change the sign of this term. From the 
Hamiltonian we can find the equations of motion for the relevant operators. 
Assuming most of the population to be in the ground state the equations of 
motion are given by
\begin{eqnarray*}
\fl
\left(\frac{\partial}{\partial t}+c\frac{\partial}{\partial z}\right)
\Eslow(z,t)
=\rmi\frac{Ng_0 \wp}{\mathcal{N}_\mathcal{S}\hbar}
\int_{-\infty}^\infty \int_{-\infty}^\infty
\mathcal{G}_0(\Delta_0)\mathcal{G}(\Delta)
\hat{\sigma}(z,t,\Delta_0, \Delta)\rmd \Delta_0 \rmd\Delta\\
\fl
\frac{\partial}{\partial t}\hat{\sigma}(z,t,\Delta_0,\Delta)
=-\rmi(\Delta_0+\Delta)\hat{\sigma}(z,t,\Delta_0,\Delta)
+\rmi\frac{\mathcal{N}_\mathcal{S}g_0 \wp}{\hbar}\Eslow(z,t)
\end{eqnarray*}
Since these equations are linear and only couple effective annihilation 
operators we can ignore the hats on the operators and consider them to be 
equations of functions instead of operators. This allows us to calculate any 
normally ordered product of operators, and evaluate the efficiency of the 
memory. If we only consider a single incoming mode as we will do throughout 
this article, this efficiency is the only important parameter for 
characterizing the performance of the memory \cite{hammerer,gorshkov2}.

Using the bandwidth of the memory $\mu$ defined above we can change to 
dimensionless units. We introduce the spatial coordinate $\Rz=z/L$ and the time 
$\Rt=\mu(t-z/c)$. We also introduce the dimensionless detunings 
$\RDelta_0 = \Delta_0/\mu$ and $\RDelta = \Delta/\mu$. Then we can define the
dimensionless broadening distributions $G_0$ and $G$ and the dimensionless 
electric field $E$. These are related to the old quantities by 
$G_0(\RDelta_0)=\mu\mathcal{G}_0(\mu\RDelta_0)$, 
$G(\RDelta)=\mu\mathcal{G}(\mu\RDelta)$ and 
$E(z,t)=\mathcal{E}_{\rm slow}(z,t)/\sqrt{\mu}$. Using these dimensionless variables the equations of 
motion become
\begin{eqnarray}
\label{E_slow_eqn_broadened}
\frac{\partial}{\partial \Rz}E(\Rz,\Rt)
=\rmi\int_{-\infty}^{\infty} \int_{-\infty}^{\infty}
G_0(\RDelta_0)G(\RDelta) \sigma(\Rz,\Rt,\RDelta_0,\RDelta)
\rmd\RDelta_0 \rmd\RDelta,\\
\label{sigma_slow_eqn}
\frac{\partial}{\partial \Rt}\sigma(\Rz,\Rt,\RDelta_0,\RDelta)
=-\rmi(\RDelta_0+\RDelta)\sigma(\Rz,\Rt,\RDelta_0,\RDelta)
+\rmi E(\Rz,\Rt).
\end{eqnarray}
To describe reversal of the controlled broadening we can replace $\RDelta$ by 
$-\RDelta$ in \eref{E_slow_eqn_broadened} and \eref{sigma_slow_eqn}. In the 
stages without any broadening (1, 3 and 5), we omit the term with $\RDelta$. 

To simplify the discussions below it will be convenient to introduce the 
polarization $P$ for each value of the intrinsic detuning. This is defined by
\begin{equation}\label{P_defined_in_terms_of_sigma}
P(\Rz,\Rt,\RDelta_0)
=\int_{-\infty}^\infty G(\RDelta)\sigma(\Rz,\Rt,\RDelta_0,\RDelta)\rmd\RDelta.
\end{equation} 
In stage 1 and 5 equations of motion \eref{E_slow_eqn_broadened} and 
\eref{sigma_slow_eqn} can be written in terms of $P$ instead of $\sigma$ since 
the controlled broadening is not present (see \eref{E_slow_eqn_stage1} and 
\eref{P_slow_stage1} in \ref{appendix_details}).

We also note that in the numerical simulations we shall assume that the 
broadening distributions are Gaussian, i.e.
\begin{equation}\label{Gaussian_distribution_definition}
G_0(\RDelta_0)=\frac{1}{\sqrt{2\pi\Rgamma_0^2}}
\exp\left(-\frac{\RDelta_0^2}{2\Rgamma_0^2}\right),
\qquad
G(\RDelta)=\frac{1}{\sqrt{2\pi\Rgamma^2}}
\exp\left(-\frac{\RDelta^2}{2\Rgamma^2}\right).
\end{equation}
Here the dimensionless widths are $\Rgamma_0 = \gamma_0/\mu$ and 
$\Rgamma = \gamma/\mu$. 

\section{Analytical theory}
\label{analytical}
Before looking at the results of the numerical simulations which provide a full 
assessment of the efficiency of the memory we can gain some intuition about 
the important aspects of the proposed protocol by doing simple perturbative 
calculations. Here we therefore investigate the process in stage~2 and 4 where 
the broadening first rapidly turns off and later turns on the 
absorption-reemission process in such a way that a large fraction of the 
excitation is left in the atoms during the storage interval (stage 3). To 
focus on this part of the dynamics we completely ignore the evolution in 
stages 1, 3 and 5 for now.

For simplicity we shall neglect the intrinsic broadening so that our system is 
described by the equations
\begin{eqnarray}
\label{E_slow_eqn_broadened_no_G0}
\frac{\partial}{\partial \Rz}E(\Rz,\Rt)=\rmi\int_{-\infty}^{\infty}
G(\RDelta) \sigma(\Rz,\Rt,\RDelta)
\rmd\RDelta,\\
\label{sigma_slow_eqn_no_G0}
\frac{\partial}{\partial \Rt}\sigma(\Rz,\Rt,\RDelta)
=-\rmi\RDelta\sigma(\Rz,\Rt,\RDelta)
+\rmi E(\Rz,\Rt).
\end{eqnarray}
We assume that all of the input pulse was mapped to $P$ in stage 1 so that 
$E(\Rz=0,\Rt)=0$. At the beginning of stage~2 the initial condition is then 
$\sigma(\Rz,\Rt=0,\Delta)=P^{(1)}(\Rz)$, where $P^{(1)}$ denotes the 
polarization at the end of stage 1. To proceed we Laplace transform
 the equations ($\Rz\rightarrow u$) and combine them into
\begin{equation*}
\frac{\partial}{\partial \Rt}\bar{\sigma}(u,\Rt,\RDelta)
=-\rmi\RDelta\bar{\sigma}(u,\Rt,\RDelta)
-\frac{1}{u} \int_{-\infty}^{\infty} G(\RDelta') \bar{\sigma}(u,\Rt,\RDelta')
\rmd \RDelta'.
\end{equation*}
We want to solve this equation perturbatively. To do this we 
introduce 
$\sigma_S(\Rz,\Rt,\RDelta)=\rme^{\rmi\RDelta\Rt}\sigma(\Rz,\Rt,\RDelta)$ which 
satisfies the differential equation
\begin{equation}\label{sigma_S_analytic_evolution_eq}
\frac{\partial}{\partial \Rt}\bar{\sigma}_S(u,\Rt,\RDelta)
=-\frac{1}{u}\rme^{\rmi\RDelta\Rt}\int_{-\infty}^\infty 
G(\RDelta')\bar{\sigma}_S(u,\Rt,\RDelta')\rme^{-\rmi\RDelta'\Rt}\rmd\RDelta'.
\end{equation}
Under the assumption that $\sigma_S$ changes slowly as function of $\Rt$ we 
can replace $\bar{\sigma}_S(u,\Rt,\Delta')$ in the integrand on the right hand 
side of \eref{sigma_S_analytic_evolution_eq} by 
$\bar{\sigma}_S(u,\Rt=0,\Delta')=\bar{P}^{(1)}(u)$. Defining the Fourier 
transform of $G$ by
$\tilde{G}(\Rt)=\int_{-\infty}^\infty G(\RDelta')
\rme^{-\rmi\RDelta'\Rt} \rmd\RDelta'$ 
the approximate solution for $\sigma$ at the end of stage~2 is
\begin{equation}\label{sigma_S_analytic_evolution_eq_solution}
\bar{\sigma}^{(2)}(u,\RDelta)
=\rme^{-\rmi\RDelta\Rtaud}
\left(1-\frac{1}{u}\int_0^{\Rtaud}
\rme^{\rmi\RDelta \Rti'}\tilde{G}(\Rti')\rmd \Rti'\right)
\bar{P}^{(1)}(u)
\end{equation}
For the case of Gaussian $G$ we have 
$\tilde{G}(\Rt)=\exp\left(-\Rgamma^2\Rt^2/2\right)$ describing the dephasing 
in time.

The expression \eref{sigma_S_analytic_evolution_eq_solution} gives the 
perturbative approximation for the evolution during stage~2 in the absence of 
an incoming pulse. In stage 4 we make the same approximation. The equations 
are the same as above but with $\RDelta$ replaced by $-\RDelta$ throughout to 
describe reversal of the broadening. Hence we actually look at the evolution 
of the polarization with reversed sign of $\RDelta$ which evolves from the 
initial condition \eref{sigma_S_analytic_evolution_eq_solution}. At the end of 
stage 4 the solution for $\sigma$ to first order in the perturbation is then
\begin{equation*}
\fl
\bar{\sigma}^{(4)}(u,-\RDelta)
=\left(1
-\frac{1}{u}\int_0^{\Rtaud} \rme^{\rmi\RDelta\Rti'}\tilde{G}(\Rti')\rmd\Rti'
-\frac{1}{u}\int_0^{\Rtaud} \rme^{\rmi\RDelta(\Rti'-\Rtaud)}
\tilde{G}(\Rti'-\Rtaud)\rmd\Rti'\right)
\bar{P}^{(1)}(u).
\end{equation*}
From this expression and the definition \eref{P_defined_in_terms_of_sigma} we 
can find $P$. Using the assumption that $G$ and hence $\tilde{G}$ is an even 
function of its argument the total polarisation at the end of stage 4 can be 
written
\begin{equation}\label{P_analytic_solution}
\bar{P}^{(4)}(u)
=\left(1-\frac{2}{u}\int_0^{\Rtaud} 
\left(\tilde{G}(\Rti')\right)^2 \rmd\Rti'\right)
\bar{P}^{(1)}(u).
\end{equation}

The main idea behind the memory protocol we propose here is that excitations 
mapped into the memory during stage 1 are released during stage 5. The 
expression above describes how an initial excitation stored in the 
polarization of the atoms described by $P^{(1)}$ at the end of stage 1 is 
mapped to the end of stage 4. To get an idea about the performance of the 
broadening mechanism we consider the efficiency with which this mapping is 
achieved. To do this we calculate the efficiency which can be expressed by
$\eta=\int_0^1 |P^{(4)}(\Rz)|^2 \rmd\Rz$ if we assume that $P^{(1)}$ is 
normalized to unity ($\int |P^{(1)}(\Rz)|^2 \rmd \Rz=1$). Now we take the 
inverse Laplace transform ($u\rightarrow\Rz$) of \eref{P_analytic_solution}. 
Assuming $G$ to be Gaussian we can carry out the integration explicitly. 
Keeping only terms to lowest order in the perturbation we find
\begin{equation}\label{P_analytic_efficiency}
\eta(\taud)
=1-2\sqrt{\pi}\frac{\mu}{\gamma}\erf(\gamma\taud)
\int_0^1 P^{(1)}(\Rz)\int_0^\Rz P^{(1)}(z')\rmd z'\rmd\Rz.
\end{equation}
This expression is the main result of this section. Ideally we would like to 
have an efficiency of unity. From the expression we see that an efficient 
memory operation can be achieved if $\gamma\gg \mu$, i.e., if the width of the 
broadening is much larger than the bandwidth of the memory. Alternatively this 
can be expressed in terms of the optical depth: as derived in 
\ref{appendix_optical_depth} the measurable optical depth in the presence of 
broadening is $d\approx\sqrt{2\pi}\mu/\gamma$. Hence the limit where the 
broadening is efficient at storing the excitation, is equivalent to the limit 
where the optical depth is much smaller than unity after the broadening has 
been turned on.

The argument of the error function in \eref{P_analytic_efficiency} reflects 
the remaining reemission which has not been completely dephased at the end of 
stage~2 (the emission rate at the end is $\sim d\eta/d\taud$). Hence if 
$\taud\gg 1/\gamma$ the reemission is completely turned off and additional 
dephasing does not improve the efficiency considerably. Therefore there is no 
need to have the controlled broadening turned on during stage 3 as we said 
earlier.

To have an idea about the validity of \eref{P_analytic_efficiency}, we can 
compare it to a numerical solution of \eref{E_slow_eqn_broadened_no_G0} and 
\eref{sigma_slow_eqn_no_G0} using the method described in 
\ref{appendix_details}. In 
figure~\ref{broadening_stages_efficiency_numeric_vs_perturbative} we plot the 
efficiencies calculated by the numerical and the perturbative approach. The 
two approaches are seen to be agree rather well for $\gamma\gg \mu$.

The results of \cite{gorshkov2} show that ideal performance of stages 1 and 5 
is achievable for sufficiently dense samples. The results obtained in this 
section demonstrate that also stage 2 and 4 can be made to work, thus 
demonstrating that an efficient memory is achievable if we can apply a strong 
reversible broadening of the atomic transition $\gamma\gg \mu$. To reach this 
result we have for simplicity assumed that the incoming field is only incident 
in stage 1. As mentioned above we, however, allow for an incoming field 
also in stage~2 of the protocol in our numerical evaluation of the efficiency 
of the protocol. As we shall see below, allowing for this small tail of the 
incoming pulse to leak into stage~2 improves the efficiency of the protocol 
and allows for a much more rapid convergence with the broadening $\gamma$ than 
predicted by the results of this section.

\begin{figure}[t]
\begin{center}
\includegraphics{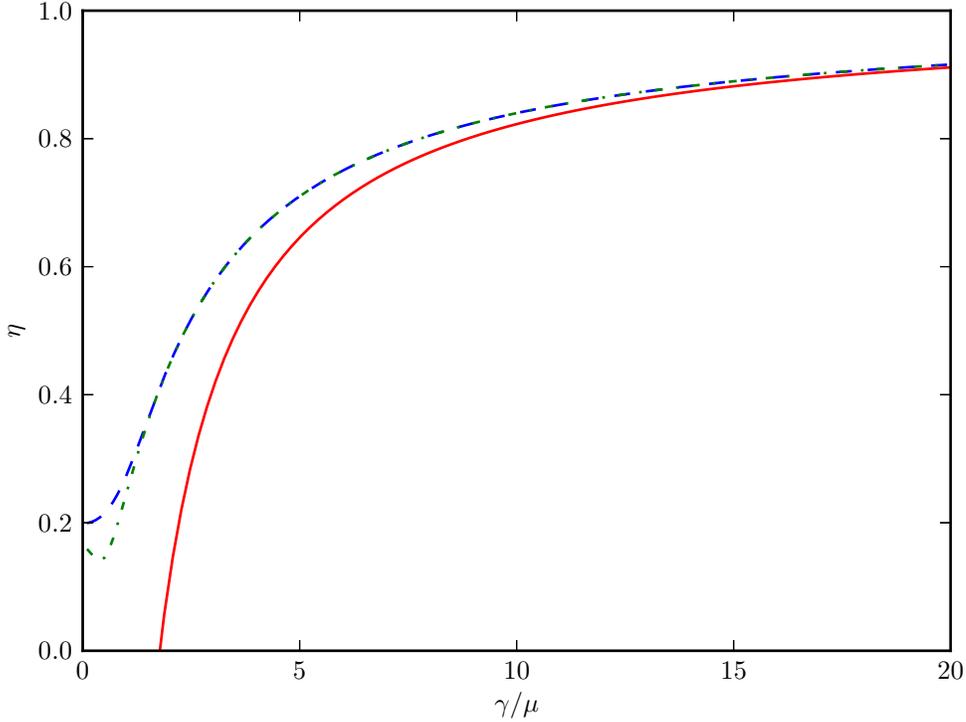}
\end{center}
\caption{Efficiencies of the broadening stages (stage~2 and stage 4) 
calculated numerically (dashed and dash-dotted lines) and perturbatively using 
\eref{P_analytic_efficiency} (solid line). In the figure we show
the efficiency for an excitation stored at the end at stage 1 to still be 
present at the end of stage 4 as a function of the broadening $\gamma$ 
relative to the bandwidth of the memory $\mu$. In all cases we assume that the 
stored polarization has the form $P^{(1)}(\Rz)=1$ which represents the worst 
case scenario for the efficiency. In the figure we also compare two different 
durations of the broadening stages $\taud=1/\mu$ (dashed) and $\taud=2/\mu$ 
(dashed-dotted). The duration of the broadening stage has little influence 
on the efficiency as long as $\gamma\taud\gg 1$. Hence in the perturbative 
case we simply take the limit $\Rtaud\rightarrow\infty$.}
\label{broadening_stages_efficiency_numeric_vs_perturbative}
\end{figure}

\section{Results}
Having verified that the memory can in principle work with near 100\% 
efficiency in the ideal limit of dense samples and large broadening, we now 
turn to a detailed numerical investigation of the performance of the protocol 
for real parameters. Specifically we shall investigate the performance of the 
memory protocol for finite optical depth and finite broadening of the atomic 
transition.

The details of the numerical procedure used to solve the equations of motion 
numerically are given in \ref{appendix_details}.
We want to find the relation between the incoming and outgoing light fields. 
Since the underlying equations are all linear ``beam-splitter equations'', 
which couple the annihilation operators of the effective harmonic oscillators, 
the relation between the incoming fields is also a beam splitter relation. As 
a result the connection between the incoming fields in stage 1 and 2 and the 
outgoing fields in stage 4 and 5 can be written in the form
\begin{equation}\label{E_out_combined}
E_{\rm out}(\Rt)
=\int_0^{\RtauR} K_E(\Rt,\Rti')
E_{\rm in}(\RtauR-\Rti')\rmd\Rti',
\end{equation}
where $\tauR=\taup+\taud$ is the total duration of read-in sequence and also 
the duration of the read-out sequence. The detailed expression for the matrix 
kernel $K_E$ can be found in \ref{appendix_details}. In the end we are 
interested in the efficiency of the whole process which is given by
\begin{equation}\label{efficiency_expression}
\eta=\int_0^{\RtauR} |E_{\rm out}(\Rt)|^2 \rmd\Rt
\end{equation}
if we assume that the incoming pulse is normalized, i.e. 
$\int_0^{\RtauR} |E_{\rm in}(\Rt)|^2 d\Rt = 1$. Using expression 
\eref{E_out_combined} we can write the efficiency as
\begin{equation}
\label{etafinal}
\eta
=\int_0^{\RtauR}\int_0^{\RtauR}
E^*_{\rm in}(\RtauR-\Rti')
K_{\rm eff}(\Rti',\Rti'')
E_{\rm in}(\RtauR-\Rti'')
\rmd\Rti' \rmd\Rti'',
\end{equation}
where
\begin{equation}\label{K_eff_definition}
K_{\rm eff}(\Rti',\Rti'')
=\int_0^{\RtauR}K_E^*(\Rt,\Rti')K_E(\Rt,\Rti'')\rmd \Rt.
\end{equation}
In the numerical simulations we discretize the time and use a quadrature rule 
\cite{Bailey06tanh-sinhhigh-precision} to setup a matrix 
$K_{\rm eff}(\Rti,\Rti')$ for the discrete times (nodes of the quadrature 
rule). With this matrix it is thus possible to obtain the memory efficiency 
for any vector containing the input fields at the discrete times.

For evaluating the performance of the memory there are various approaches that 
one can take. From \eref{etafinal} we see that the memory efficiency depends 
on the shape of the incoming pulse. One approach to evaluating the performance 
is to look for the incoming mode which has the highest efficiency. Since the 
final expression for the efficiency \eref{etafinal} can be written in the form 
of a simple vector and matrix product, where the kernel matrix $K_{\rm eff}$ 
is self-adjoint, the maximal efficiency can be shown to be given by the 
largest eigenvalue of the matrix $K_{\rm eff}$. Alternatively the performance 
of the memory can be assessed by investigating how well the memory operates 
with a specially chosen mode that one may be interested in. Below we shall 
consider both these approaches. We emphasize, however, that there are many 
more methods of characterizing the performance. For instance the approaches 
that we take here do not characterize the ability of the memory to store 
multiple modes \cite{nunnmultimode} and also ignore any 
information about the shape of the outgoing pulse, which may important for 
practical applications.

To simulate the performance of the memory we consider a realistic situation 
where we have a collection of atoms with an intrinsic broadening of the 
optical transition described by the Gaussian distribution $G_0$ defined in 
\eref{Gaussian_distribution_definition} with width $\Rgamma_0=\gamma_0/\mu$. 
Since we are interested in a quantum memory operating only on two-level atoms 
the width of this distribution inherently leads to a decay of the atomic 
polarization by an amount $\exp(-(\taus/T_2)^2)$ with $T_2=d_0/(\sqrt{\pi}\mu)$ 
as shown in \ref{appendix_optical_depth}. At the same time, the width of the 
atomic line also determines the measurable optical depth $d_0$ of the ensemble 
before the controlled broadening is turned on which is given by 
$d_0=\sqrt{2\pi}\mu/\gamma_0$. In the investigations below we are mainly 
interested in how the performance scales with the optical depth of the atomic 
ensembles. We therefore fix the storage time $\taus$ to be equal to 
$1/2\gamma_0=T_2/\sqrt{8}$ so that the maximal attainable storage efficiency 
is equal to $\exp(-\taus^2\gamma_0^2)=\exp(-1/4)\approx 0.78$. With the 
storage time fixed relative to the dephasing time $T_2$, the investigations of 
the dependence of the optical depth below essentially correspond to the 
scaling one obtains when varying the number of atoms in the ensemble while 
keeping all other parameters fixed. The allowed duration of the pulse to be 
stored is primarily determined by the duration $\taup$ of stage 1. For a 
memory to make sense the duration of the pulse must be shorter than the memory 
time $\taup<\taus$. We therefore restrict ourselves to a duration 
$\taup=\taus/4$. Finally we chose a constant duration $\taud$ of stages 2 and 4. 
As shown above, the duration of this period is not so important as long as it 
is long enough that after stage~2, all the different polarizations are 
dephased sufficiently so that that negligible light gets out during stage 3. 
In the numerical simulations we are, however, constrained by having only a 
finite number of discrete frequencies. This means that there exists a finite 
time when the polarizations rephase again. Hence $\taud$ has to be chosen such 
that it is well below this rephasing time and we fix it at $\taud=1/\mu$.

\begin{figure}[hbt]
\begin{center}
\includegraphics{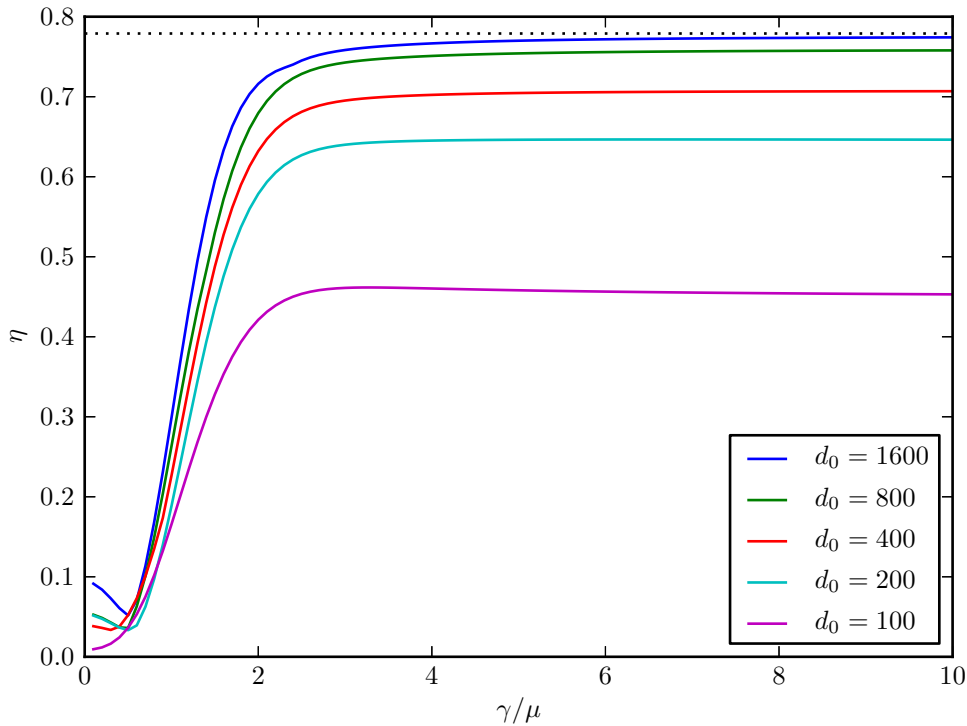}
\end{center}
\caption{Efficiency $\eta$ for the optimal incoming mode and a Gaussian 
distribution of the controlled broadening. For a given set of parameters 
characterizing the memory we find the mode which gives the highest possible 
storage and retrieval efficiency and plot it as a function of the width of the 
controlled broadening $\gamma/\mu$ for different optical depths $d_0$. Here 
$d_0$ is the optical depth before the broadening is applied. Values of $d_0$ 
in the legend are given in the same order as values of $\eta$ at 
$\gamma/\mu=10$. In the simulation the storage time $\taus$ is fixed at 
$\taus=T_2/\sqrt{8}$, where $T_2$ is the dephasing time. This limits the 
efficiency to $\eta\leq \exp(-2 (\taus/T_2)^2)\approx 0.78$, which is shown as 
a dotted curve. For large optical depths $d_0\gg 1$ and large applied 
broadening $\gamma\gtrsim 3/\mu$ the curves approach the upper limit showing 
that we can have an efficient memory. The parameter $\mu=d_0/(\sqrt{\pi}T_2)$ 
characterizes the bandwidth of the memory and when the broadening is applied 
it reduces the optical depth to a value $d\approx \sqrt{2\pi}\mu/\gamma$. In 
the simulation we allow for an incoming pulse of duration $\taup=\taus/4$ and 
the dephasing is applied for a time $\taud=1/\mu$.}
\label{optimal_mode_efficiency}
\end{figure}

In figure~\ref{optimal_mode_efficiency} we show the maximal efficiency 
obtainable for a given optical depth and width of the broadening. In the 
figure we see that the efficiency rapidly increases when we apply the 
broadening and saturates when the width of the applied broadening reaches a 
value around $\gamma\gtrsim 3\mu$. Once the applied broadening reaches this 
value it is sufficiently broad to rapidly dephase the polarization and thus 
rapidly turn off the reemission of the absorbed light after the broadening has 
been applied. Furthermore we see that as we increase the optical depth of the 
ensemble $d_0$ the efficiency approaches the maximally allowed efficiency of 
$\eta\approx 0.78$. The procedure proposed here thus allows for efficient 
quantum memory operation using only two-level atoms. In particular by allowing 
the broadening to be turned on after the pulse is incident we are able to 
surpass the limit of $\eta=0.54$ identified in \cite{crib} for a two-level 
quantum memory based on the standard transverse CRIB approach where the 
broadening is turned on before the pulse enters the medium 
(figure~\ref{principle_of_operation_diagram}~a). Furthermore for a sample with 
a long coherence time $T_2\gg \tau_s$ and a high optical depth $d\gg 1$ we can 
in principle come arbitrarily close to an efficiency of 100\%.

\begin{figure}[hbt]
\begin{center}
\subfigure[]{\label{optimal_mode_plot1}
\includegraphics{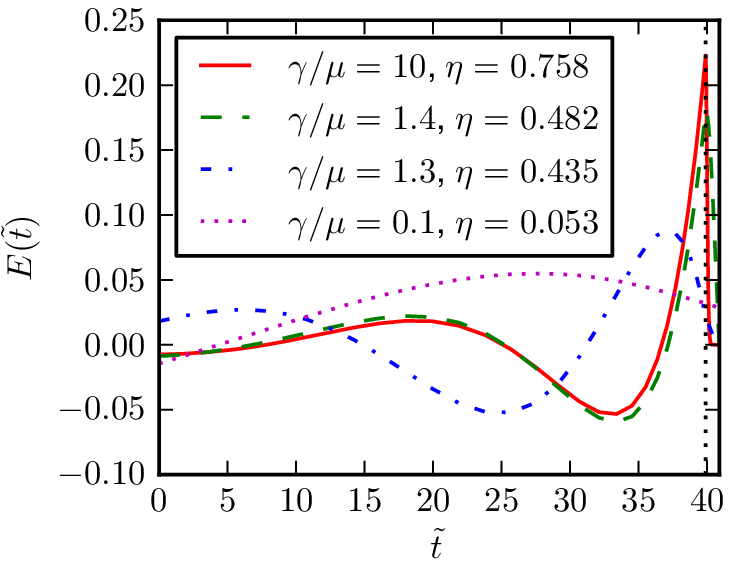}}
\subfigure[]{\label{optimal_mode_plot2}
\includegraphics{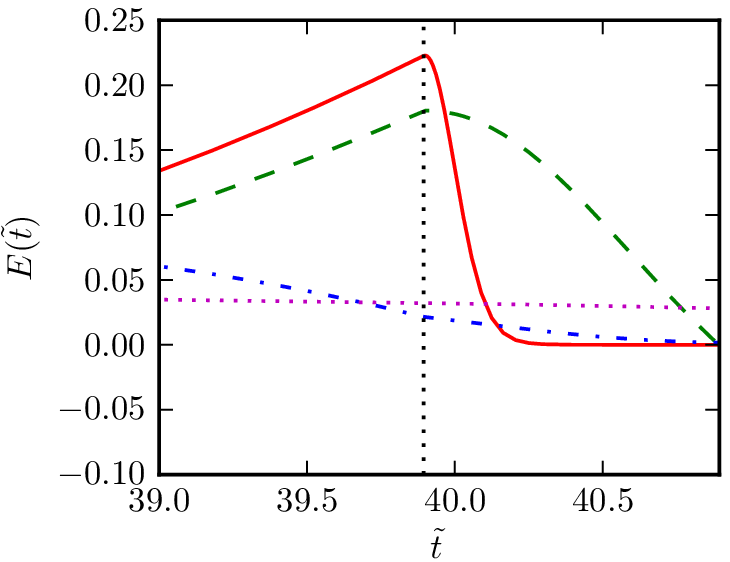}}
\end{center}
\caption{The optimal mode shapes corresponding to the maximal efficiencies in 
figure~\ref{optimal_mode_efficiency} with $d_0=800$. Here the controlled 
broadening is applied at $\Rt=\mu\taup=d_0/(8\sqrt{2\pi})\approx 39.89$, which 
is shown as a vertical dotted line. (a) optimal mode shape for different widths 
of the applied broadening. For $\gamma\gtrsim 1.4 \mu$ the mode shape 
approaches the Bessel-like modes identified in \cite{gorshkov2}. (b) zoom in 
of (a) around $\Rtaup$.}
\label{optimal_mode_plot}
\end{figure}

Some examples of the optimal modes leading to the maximal efficiency in 
figure~\ref{optimal_mode_efficiency} are shown in 
figure~\ref{optimal_mode_plot}. For low values of the broadening, where the 
memory is inefficient, the optimal shape is a rather flat. As we begin to 
increase the width of the applied broadening the optimal shape changes 
character. After a certain value of $\gamma$ which in this case is 
$\gamma\approx 1.4\mu $ the shape of the modes for $\Rt\leq\Rtaup$ begins to 
resemble a Bessel like function, which is the shape of optimal modes 
identified in \cite{gorshkov2} (plotted in \cite{gorshkov3}) for the so called 
fast memory regime, which corresponds to our stage 1. This resemblance 
reflects that the mechanism in these approaches are highly similar. The only 
difference is that our dephasing mechanism which shuts off the reemission of 
the stored field is replaced in \cite{gorshkov2} by a $\pi$-pulse taking the 
excitation from $|e\rangle$ to an auxiliary state. After the broadening is 
applied at $\Rt=\Rtaup$ the optimal mode shape rapidly drops to zero on a time 
scale set by the width of the broadening. This reflects that $1/\gamma$ is the 
time scale needed dephase the excitation and thus terminating the read-in 
sequence.

It is interesting to note that the efficiency in 
figure~\ref{optimal_mode_efficiency} approaches the stationary value much more 
rapidly when we increase the width $\gamma/\mu$ than shown in 
figure~\ref{broadening_stages_efficiency_numeric_vs_perturbative}. The reason 
for this difference is that we allow for a small tail of the pulse to leak 
into stage~2, which was not included in section \ref{analytical}. In stage~2 
the effective optical depth is given by 
$d=\sqrt{2\pi}\mu/\sqrt{\gamma^2+\gamma_0^2}\approx \sqrt{2\pi}\mu/\gamma$ for 
$\gamma\gtrsim \gamma_0=\sqrt{2\pi}\mu/ d_0$. Hence for the regime where the 
memory is efficient $\gamma >\mu$ the optical depth is below unity during 
stage~2. Regardless of this, the inclusion of a small optical field during 
this stage is still sufficient to alter the memory efficiency. This result 
emphasizes that the optical depth is not necessarily the correct physical 
parameter for characterizing the transient absorption of a pulse. The optical 
depth characterizes the fraction of the incident energy which is absorbed and 
not reemitted if the parameters of the memory are stationary. In our dynamic 
situation where the parameters are varying in time the optical depth does not 
correctly characterize the transient absorption of the pulse. In fact in our 
protocol most of the energy of the pulse is absorbed during stage 1 where the 
atomic line width is very narrow. Hence the pulse that we store is much 
broader in frequency than the atomic line width and most frequency components 
of the field see an optical depth much less than unity. This, however, only 
means that there is no stationary absorption and does not exclude that the 
field is absorbed and reemitted several times during the passage through the 
memory (e.g. the reduced group velocity of a pulse traveling through a 
transparent medium, such as glass, can be understood as a consequence of 
constant absorption and reemission events for the light. In this case the 
resulting pulse delay can be substantial even if the optical depth is very 
low). In this sense the idea behind the current memory protocol is to 
interrupt the frequent absorption and reemission events by applying the 
broadening which stops and later resumes this absorption and reemission 
process.

\begin{figure}[hbt]
\begin{center}
\subfigure[$\gamma=0.1\mu$]{\label{gaussian_mode_efficiency_contour1}
\includegraphics{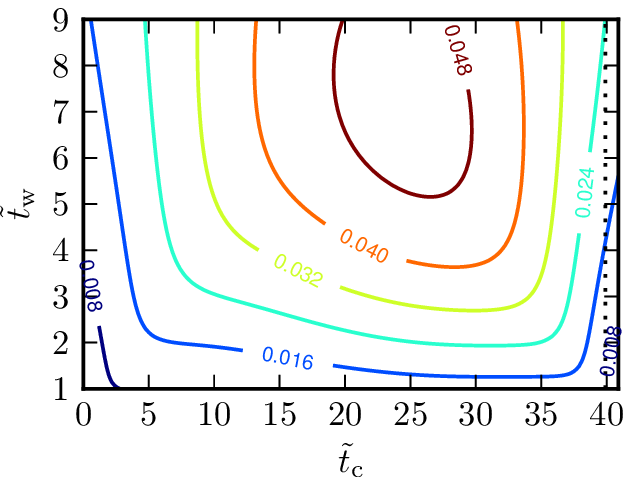}}
\subfigure[$\gamma=\mu$]{\label{gaussian_mode_efficiency_contour2}
\includegraphics{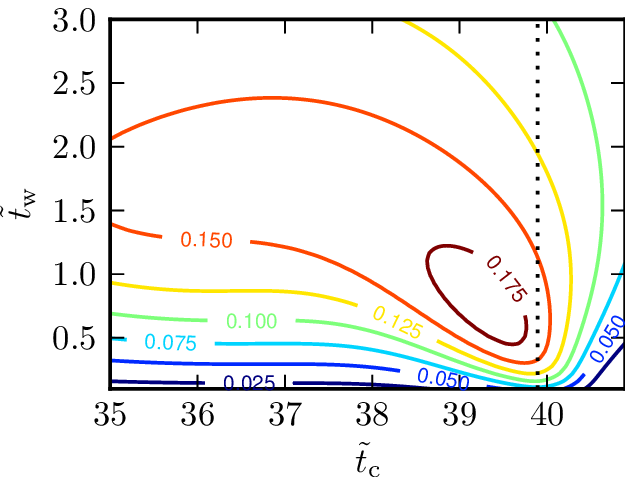}}
\subfigure[$\gamma=10\mu$]{\label{gaussian_mode_efficiency_contour3}
\includegraphics{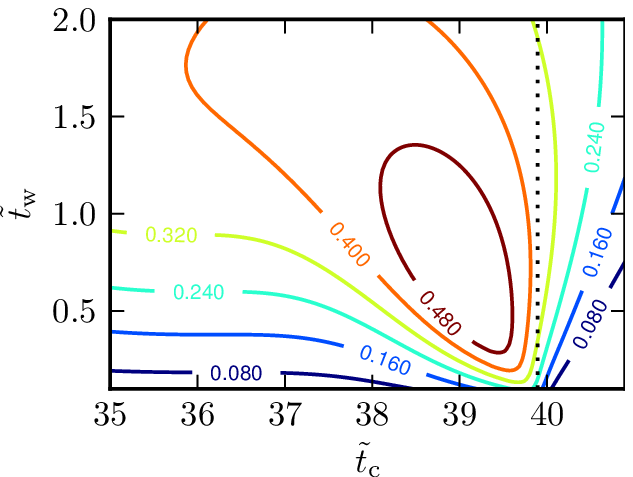}}
\end{center}
\caption{Efficiency of Gaussian modes for $d_0=800$ and different widths 
$\Rt_{\rm w}$ and center $\Rt_{\rm c}$ of the Gaussian mode. The three 
different plots (a, b, and c) show the efficiencies for various different 
widths of the applied broadenings. Here $\Rt_{\rm c}=\mu\taup\approx 39.89$ is 
the time at which the applied broadening is turned on (shown as vertical 
dotted line). For $\gamma\gg \mu$ the best performance is achieved for narrow 
pulses which are incident shortly before the broadening is applied.}
\label{gaussian_mode_efficiency_contour}
\end{figure}

Above we have focused on the maximal efficiency obtainable for the optimal 
mode shape. Since these optimal modes may not be available experimentally we 
can also consider how the memory performs for a given predefined mode shape. 
We therefore also calculate the efficiency for Gaussian input modes of the 
form
\begin{equation}\label{gaussian_mode_E_in}
E_{\rm in}(\Rt, \Rt_{\rm c}, \Rt_{\rm w})
=\frac{1}{(2\pi\Rt_{\rm w}^2)^{1/4}}
\exp\left(-\frac{(\Rt-\Rt_{\rm c})^2}{4\Rt_{\rm w}^2}\right).
\end{equation}
This mode is characterized by two parameters, the center $\Rt_{\rm c}$ and and 
the width $\Rt_{\rm w}$, and in figure~\ref{gaussian_mode_efficiency_contour} 
we show the efficiency as a function of these parameters. As we can see in the 
figure there is always a well defined maximum that we can find via numerical 
optimization. Furthermore in the limit where we expect the memory to work 
$\gamma\gg\mu$ the best performance is achieved by sending in a rather narrow 
pulse of duration $\Rt_{\rm w}\sim \mu$ right before broadening is applied. 
This results indicates that the best performance is achieved when the incident 
pulse resembles the sharply peaked optimal mode functions in 
figure~\ref{optimal_mode_plot}. 

\begin{figure}[hbt]
\begin{center}
\includegraphics{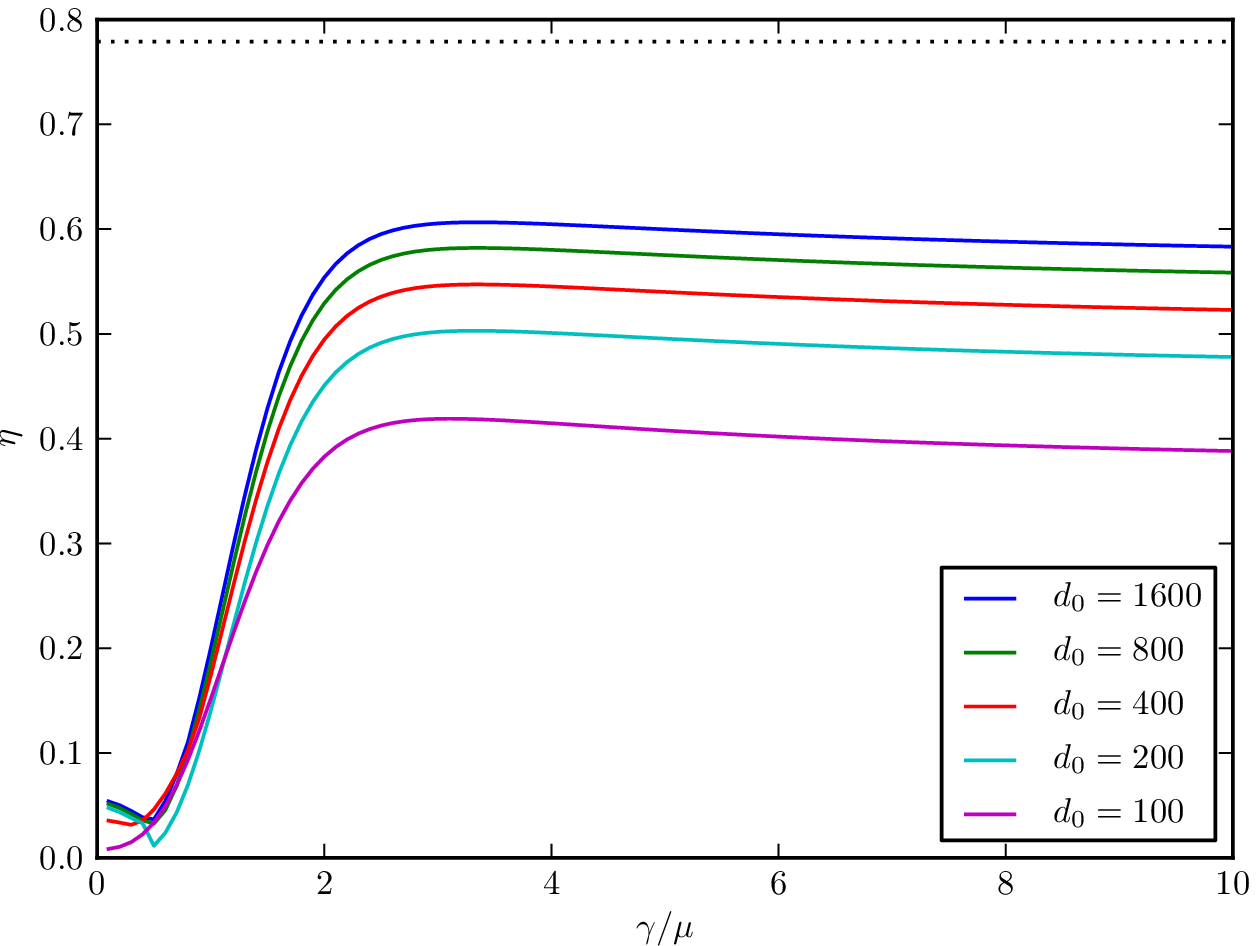}
\end{center}
\caption{Maximal efficiency of Gaussian modes \eref{gaussian_mode_E_in} for 
different optical depths before the broadening is applied $d_0$. Values of 
$d_0$ in the legend are given in the same order as values of $\eta$ at 
$\gamma/\mu=10$. The efficiency of a Gaussian pulse is optimized with respect 
to the width and center of the pulse and the figure shows the optimal 
efficiency. Similar to figure~\ref{optimal_mode_efficiency} the efficiency 
increases with the broadening and reaches the maximal value for 
$\gamma\gtrsim 3\mu$. The efficiency also increases with increasing optical 
depth but the increase is much slower than for the optimal modes in figure 
\ref{optimal_mode_efficiency}. The dotted line at $\eta\approx0.78$ represents 
the maximal attainable efficiency due to the dephasing during the memory 
time.}
\label{best_gaussian_mode_efficiency}
\end{figure}

If we optimize $\Rt_{\rm w}$ and $\Rt_{\rm c}$ 
we can find the optimal Gaussian pulse for storage into the memory.
The results of this optimization are shown in 
figure~\ref{best_gaussian_mode_efficiency}. Here we see that for the optimal 
Gaussian mode the efficiency also rapidly increases with the applied 
broadening and reaches a maximum for $\gamma\approx 3\mu$. Then efficiency 
actually starts to decrease which is a consequence of the Gaussian mode shape not 
resembling the changing optimal mode shape. The efficiency also increases with 
increasing optical depth of the ensemble $d_0$ although much slower than for 
the optimal mode in figure~\ref{optimal_mode_efficiency}. Unfortunately 
limited numerical resources prevent us from increasing the optical depth even 
further, but we believe that it will eventually converge towards the maximal 
attainable efficiency of $\eta\leq \exp(-2 (\taus/T_2)^2)\approx 0.78$. 
Nevertheless our results are still able to surpass the limit of 
$\eta=54\% \cdot 0.78=42\%$ for the two-level memory protocol presented in 
\cite{crib} when we take into account the dephasing during the memory time.

\section{Conclusion and discussion}

We have proposed a method to make an efficient quantum memory for light based 
on two-level atoms. In our proposal the memory operation is controlled by 
applying and later reversing an external field which broadens the atomic 
transition. For sufficiently strong broadening and sufficiently high optical 
depth before the broadening is applied, we have shown that the memory 
operation can be limited only by the intrinsic decoherence (the $T_2$ time) of 
the atomic transition. Most importantly, in contrast to most protocols the 
proposed memory can be efficient even without employing any optical control 
field. Since there is no need for an additional laser field we hope that this 
memory protocol may be simpler to implement in practice.

Our protocol is not the first proposal for a quantum memory based on two-level 
atoms. In \cite{longcrib,longcribprl} an efficient memory was proposed for the 
so-called {\it longitudinal} CRIB where a field gradient is applied such that 
the shift of the atomic transition depends on the position. On the contrary 
our protocol is based on {\it transverse} CRIB where the shift does not depend 
on the position. For 
this setting the only previous protocol \cite{crib} based on two-level systems 
had a maximal efficiency of 54\%. For our protocol there is no such limit and 
the efficiency can approach 100\% for sufficiently high optical depth. The 
efficiency is, however, dependent on the mode shape and converges to the ideal 
limit much more rapidly for the optimal mode shape than if we constrain it to 
be, e.g., a Gaussian. It would be interesting to study in more detail how the 
protocol developed here compares to the other memory protocols based on two 
level systems. The previous work, however, had a different scope than what we 
consider here and only studied the dependence on the optical depth after the 
broadening was applied and the effects of the intrinsic broadening were not 
included in the final results for the efficiency. Within our framework these 
previous studies thus correspond to the limit $d_0\rightarrow\infty$ and 
cannot directly be compared to our results.

It has previously been argued that adding and reversing (transverse) 
broadening during the storage process generally reduces the memory efficiency 
for the optimal modes as compared to schemes which do not have this broadening 
\cite{gorshkov3}. In our scheme the broadening is absent during the period 
where most of the light is read into the memory. We therefore believe that for 
a given memory our proposed protocol will have the highest possible efficiency 
for the optimal mode. On the other hand it has also been shown that the 
previous CRIB protocols \cite{crib, longcribprl} have a much larger multimode
capacity than 
protocol such as this one, where the broadening is not present during the 
storage \cite{nunnmultimode}. This limited multimode capacity of our scheme is 
reflected in our results for Gaussian pulse shapes, where the efficiency 
becomes limited by the ability of the Gaussian to resemble the relatively few 
modes which are stored with high efficiency. In fact for the readout process 
we find that also for the case of an incoming Gaussian pulse, the outgoing 
mode shape resembles the outgoing mode shape for the optimal mode. Again this 
is most likely a consequence of the limited multimode efficiency: only a few 
storage modes are excited by the Gaussian, and these modes are later retrieved 
into outgoing modes which contain little information about the incoming pulse 
shape. This reshaping of the mode may be detrimental for some applications of 
the quantum memory, but for other applications the mode shape may be less
important. For instance for quantum repeaters 
\cite{dlczrepeater,reviewrepeater} we are interested in interfering the output 
from two different quantum memories. Since both memories in this case will 
emit similar mode shapes, the reshaping of the mode is of minor consequence. 
For concrete applications of quantum memories a more detailed study will have 
to be performed to determine whether the increased efficiency of the present 
protocol outweighs the drawback of the limited multimode capacity.

\section{Acknowledgements}
We thank A. Grodecka-Grad and N. Sangouard for helpful discussions. This work 
was supported by the Danish National Research Foundation and by the European 
Research Council under the European Union's Seventh Framework Programme 
(FP/2007-2013) / ERC Grant Agreement n. 306576'

\appendix
\section{Relation of the parameters of the model to physically measurable 
quantities}\label{appendix_optical_depth} In this section we relate the 
parameters of the model to physically measurable quantities. Specifically we 
shall express everything in terms of the optical depth in the absence of 
controlled broadening $d_0$ and the coherence time of the polarization of the 
atoms $T_2$. To do this we consider light propagating through the atomic 
ensemble without the controlled broadening being present. This situation can 
be described by \eref{E_slow_eqn_stage1} and \eref{P_slow_stage1}. If we 
formally integrate \eref{P_slow_stage1} under the initial condition 
$P(\Rz, \Rt=-\infty,\RDelta_0)=0$ we get
\begin{equation*}
P(\Rz,\Rt,\RDelta_0)
=\rmi\int_{-\infty}^\Rt\exp(-i\Delta_0(\Rt-\Rti'))E(\Rz, \Rti')\rmd \Rti'
\end{equation*}
Taking the Fourier transform ($\Rt\rightarrow\omega$) this equation becomes
\begin{equation}\label{P_slow_stage1_formal_solution_ft}
P(\Rz,\omega,\RDelta_0)
=\rmi\tilde{E}(\Rz, \omega)
\left(\pi\delta(\omega-\RDelta_0)-\PV\frac{\rmi}{\omega-\RDelta_0}\right)
\end{equation}
where $\PV$ reminds that we need to take the Cauchy principal value when 
integrating. Inserting \eref{P_slow_stage1_formal_solution_ft} into the 
Fourier transform of \eref{E_slow_eqn_stage1} results in
\begin{equation*}
\frac{\partial}{\partial \Rz}\tilde{E}(\Rz,\omega)
=-\tilde{E}(\Rz,\omega)\left(\pi G_0(\omega)
-\PV\int_{-\infty}^{\infty} 
G_0(\RDelta_0')\frac{\rmi}{\omega-\RDelta_0'}\rmd \RDelta_0'\right). 
\end{equation*}
From this we see that
\begin{equation*}
|\tilde{E}(\Rz=1,\omega)|^2
=\exp\left(-2\pi G_0(\omega))\right
|\tilde{E}(\Rz=0,\omega)|^2.
\end{equation*}
From this expression we define the optical depth on resonance as 
$d_0=2\pi G_0(0)$. If the controlled broadening is present this amounts to 
having a distribution of the sum of the controlled and intrinsic broadening 
$G_{\rm sum}$ instead of $G_0$. Hence we have $d=2\pi G_{\rm sum}(0)$. If both 
$G_0$ and the controlled broadening distribution $G$ are Gaussian with widths 
$\gamma_0/\mu$ and $\gamma/\mu$ respectively then $G_{\rm sum}$ is also 
Gaussian with width $\gamma_{\rm sum}/\mu=\sqrt{\gamma_0^2+\gamma^2}/\mu$. In 
the limit $\gamma \gg \gamma_0$ we simply have $\gamma_{\rm sum}\approx \gamma$, 
so that
\begin{equation*}
d_0
=\sqrt{2\pi}\frac{\mu}{\gamma_0}
\qquad
{\rm and}
\qquad
d
=\sqrt{2\pi}\frac{\mu}{\sqrt{\gamma_0^2+\gamma^2}}
\approx\sqrt{2\pi}\frac{\mu}{\gamma}.
\end{equation*}

To derive the temporal decay of the polarization we can assume that $P$ in 
\eref{P_defined_in_terms_of_sigma} is independent of $\RDelta_0$ initially, 
i.e. $P(\Rz,\Rt=0,\Delta_0)=P_0(\Rz)$. If no electric field is present, the 
evolution is given by \eref{P_slow_stage1} and the total polarization becomes
\begin{equation*}
P_{\rm total}(\Rz,\Rt)
=N\int_{-\infty}^\infty
G_0(\RDelta) P(\Rz, \Rt,\Delta_0) \rmd\RDelta
=N P_0(\Rz)\int_{-\infty}^\infty
G_0(\RDelta) \rme^{-\rmi\RDelta_0\Rt} \rmd\RDelta.
\end{equation*}
Hence for a Gaussian $G_0$ given by \eref{Gaussian_distribution_definition} we 
have
\begin{equation*}
P_{\rm total}(\Rz,\Rt)
=P_{\rm total}(\Rz,\Rt=0)\exp\left(-\frac{\Rt^2}{\mu^2 T_2^2}\right)
\end{equation*}
with $T_2=\sqrt{2}/\gamma_0=d_0/(\mu\sqrt{\pi})$.

\section{Details of the numerical method used to simulate the evolution.}
\label{appendix_details}

In this appendix we give the details of how we do the numerical simulations of 
the memory protocol. In short we Laplace transform the equations of motion in 
space ($\Rz\rightarrow u$) and rewrite them such that we obtain an equation 
only involving the polarization. Then we discretize the broadening variables 
$\RDelta_0$ and $\RDelta$ such that the integrals become sums and convolutions 
become matrix products. The resulting vector equations have simple solutions 
expressed in terms of the matrix exponential. In the end we can find the Laplace 
transform of the electric field using the discretized and Laplace transformed 
versions of \eref{E_slow_eqn_broadened} and \eref{E_slow_eqn_stage1}.
Then the electric field in real space can be evaluated by 
applying numerical inverse Laplace transform which amounts to transforming the 
integration kernels. Detailed calculations for each stage are given below.

\subsection{Stage 1}
In stage 1 the controlled broadening is not present so that $\sigma$ does
not depend on $\RDelta$. Hence using the definition 
\eref{P_defined_in_terms_of_sigma} we see that 
$\sigma(\Rz,\Rt,\RDelta_0,\RDelta)=P(\Rz,\Rt,\RDelta_0)$. Also the term 
$-\rmi\RDelta\sigma(\Rz,\Rt,\RDelta_0,\RDelta)$ is not present on the right 
hand side of \eref{sigma_slow_eqn} so that we can write it together with 
\eref{E_slow_eqn_broadened} as
\begin{eqnarray}\label{E_slow_eqn_stage1}
\frac{\partial}{\partial \Rz}E(\Rz,\Rt)=\rmi \int_{-\infty}^{\infty}
G_0(\RDelta_0) P(\Rz,\Rt,\RDelta_0)
\rmd\RDelta_0,\\
\label{P_slow_stage1}
\frac{\partial}{\partial \Rt}P(\Rz,\Rt,\RDelta_0)
=-\rmi\RDelta_0 P(\Rz,\Rt,\RDelta_0)
+\rmi E(\Rz,\Rt).
\end{eqnarray}
Now we Laplace transform these equations in space ($\Rz\rightarrow u$). 
Combining the resulting equations and using the definition 
$E_{\rm in}(\Rt)=E(\Rz=0,\Rt)$ we get
\begin{equation}\label{P_stage1_laplace_isolated}
\fl
\frac{\partial}{\partial \Rt}\bar{P}(u,\Rt,\RDelta_0)
=-\rmi\RDelta_0 \bar{P}(u,\Rt,\RDelta_0)
-\frac{1}{u}\int_{-\infty}^\infty G_0(\RDelta_0')
\bar{P}(u,\Rt,\RDelta_0')\rmd\RDelta_0'
+\frac{\rmi}{u}E_{\rm in}(\Rt).
\end{equation}
 
We choose $K$ discrete values of $\RDelta_0$ that we call
$\RDelta_{01}, \ldots, \RDelta_{0K}$. For simplicity we choose them such that 
they have a constant step $\RDelta_{0\delta}$. After discretization 
\eref{P_stage1_laplace_isolated} becomes
\begin{equation}\label{P_stage1_laplace_discrete1}
\fl
\frac{\partial}{\partial \Rt}\bar{P}(u,\Rt,\RDelta_{0j})
=-\rmi\RDelta_{0j} \bar{P}(u,\Rt,\RDelta_{0j})
-\frac{1}{u}\sum_{k=1}^K
G_0(\RDelta_{0k})\RDelta_{0\delta}P(u,\Rt,\RDelta_{0k})
+\frac{\rmi}{u}E_{\rm in}(\Rt)
\end{equation}

To simplify the notation we wish to write the discretized equations in vector 
form. To do this we define vectors $\mathbf{\Delta}_0$, $\mathbf{P}(u,\Rt)$ 
and $\mathbf{g}_0$ with elements given by $(\mathbf{\Delta}_0)_j=\RDelta_{0j}$, 
$(\mathbf{P}(u,\Rt))_j=\bar{P}(u,\Rt,\RDelta_{0j})$ and 
$(\mathbf{g}_0)_j=\RDelta_{0\delta}G_0(\RDelta_{0j})$ for $1\leq j\leq K$.
Also for any vector $\mathbf{v}$ we define a matrix $D(\mathbf{v})$ with 
elements $D(\mathbf{v})_{jj}=v_j$ for all $j$ and $D(\mathbf{v})_{jk}=0$ for 
$j\neq k$, i.e. a diagonal matrix with $\mathbf{v}$ as its diagonal. If we 
further let $\mathbf{h}^{(K)}$ be a $K$-dimensional vector with constant 
elements $h_j=1$ for all $j$ then we can write 
\eref{P_stage1_laplace_discrete1} as a vector equation
\begin{equation}\label{P_stage1_laplace_matrix1}
\fl
\frac{\partial}{\partial \Rt}\mathbf{P}(u,\Rt)
=-\rmi D(\mathbf{\Delta}_0) \mathbf{P}(u,\Rt)
-\frac{1}{u}\mathbf{h}^{(K)} \mathbf{g}_0^T\mathbf{P}(u,\Rt)
+\frac{\rmi}{u}E_{\rm in}(\Rt)\mathbf{h}^{(K)}.
\end{equation}
Note that $\mathbf{h}^{(K)} \mathbf{g}_0^T$ is a matrix with each row equal to 
the vector $\mathbf{g}_0$. Defining
\begin{equation*}
M_1(u)=-\rmi D(\mathbf{\Delta}_0)-\frac{1}{u}\mathbf{h}^{(K)} \mathbf{g}_0^T
\end{equation*}
we can write 
\eref{P_stage1_laplace_matrix1} as
\begin{equation}\label{P_stage1_laplace_matrix2}
\frac{\partial}{\partial \Rt}\mathbf{P}(u,\Rt)
=M_1(u)\mathbf{P}(u,\Rt)
+\frac{\rmi}{u}E_{\rm in}(\Rt)\mathbf{h}^{(K)}.
\end{equation}

In stage 1 the initial condition is that the memory is empty, i.e 
$P(\Rz,\Rt=0)=0$ or equivalently $\mathbf{P}(u,\Rt=0)=0$. The solution 
to \eref{P_stage1_laplace_matrix2} can then be expressed using 
the matrix exponential
\begin{equation}\label{P_stage1_laplace_matrix2_solution2}
\mathbf{P}(u,\Rt)=\rmi\int_0^\Rt \frac{1}{u}\exp(M_1(u)\Rti')
\mathbf{h}^{(K)} E_{\rm in}(\Rt-\Rti')\rmd \Rti'.
\end{equation}

\subsection{Stage 2}\label{stage2_appendix_subsection}
The memory protocol works by having most of the light absorbed during stage 1. 
Nevertheless we still allow for some light also during stage~2. Hence stage~2 
is described by \eref{E_slow_eqn_broadened} and \eref{sigma_slow_eqn} with 
the initial conditions
\begin{eqnarray*}
&E(\Rz=0,\Rt)=E_{\rm in}(\Rt+\Rtaup),\\
&\sigma(\Rz,\Rt=0,\RDelta_0,\RDelta)
=\sigma^{(1)}(\Rz,\RDelta_0,\RDelta)
=P^{(1)}(\Rz,\RDelta_0)
\end{eqnarray*}
where $\sigma^{(1)}$ and $P^{(1)}$ are two ways to denote the polarisation at 
the end of stage 1; and $E_{\rm in}$ is the same function that was defined for 
stage 1. Here and in the beginning of all the subsequent stages the time $\Rt$ 
represents the time since the beginning of the current stage. This resetting of 
time is convenient since the equations of motion change from one stage to 
another. Since we want to consider $E_{\rm in}$ as being an input pulse for 
both stage~1 and stage~2 we have to shift its argument by the duration of 
stage~1 when we solve the equations of motion in stage~2.

Taking the Laplace transform of \eref{E_slow_eqn_broadened} and 
\eref{sigma_slow_eqn} and combining the resulting equations as for stage~1 we 
get
\begin{equation}\label{sigma_stage2_laplace_isolated}
\fl
\begin{array}{*3{>{\displaystyle}l}}
\frac{\partial}{\partial \Rt}\bar{\sigma}(u,\Rt,\RDelta_0,\RDelta)
&=&-\rmi(\RDelta_0+\RDelta) \bar{\sigma}(u,\Rt,\RDelta_0,\RDelta)\\
&&-\frac{1}{u}\int_{-\infty}^\infty \int_{-\infty}^\infty
G_0(\RDelta_0') G(\RDelta')
\bar{\sigma}(u,\Rt,\RDelta_0',\RDelta)\rmd\RDelta_0'\rmd\RDelta'\\
&&+\frac{\rmi}{u}E_{\rm in}(\Rt+\Rtaup)
\end{array}
\end{equation}
Now we have to split both detunings $\RDelta_0$ and $\RDelta$ into 
respectively $K$ discrete values $\RDelta_{01}, \ldots, \RDelta_{0K}$ and $N$ 
discrete values $\RDelta_{1}, \ldots, \RDelta_{N}$. Again we assume that they 
have constant steps $\RDelta_{0\delta}$ and $\RDelta_\delta$ respectively. 
Defining $\bar{\sigma}_{j,k}(u,\Rt)=\bar{\sigma}(u,\Rt,\RDelta_{0j},\RDelta_k)$, 
the discretized version of \eref{sigma_stage2_laplace_isolated} can be written 
as
\begin{equation}\label{sigma_stage2_laplace_discrete}
\fl
\begin{array}{*3{>{\displaystyle}l}}
\frac{\partial}{\partial \Rt}\sigma_{j,k}(u,\Rt)
&=&-\rmi(\RDelta_{0j}+\RDelta_k) \sigma_{j,k}(u,\Rt)\\
&&-\frac{1}{u}\sum_{j'=1}^K \sum_{k'=1}^N
G_0(\RDelta_{0j'}) G(\RDelta_{k'})\RDelta_{0\delta}\RDelta_\delta
\sigma_{j',k'}(u,\Rt)\\
&&+\frac{\rmi}{u}E_{\rm in}(\Rt+\Rtaup)
\end{array}
\end{equation}
We see that we can write this expression in vector notation as before if we 
combine the indices $j$ and $k$ into one. Hence we define vectors 
$\mathbf{\Delta}_\pm$, $\boldsymbol\sigma(u,\Rt)$ and $\mathbf{g}$ with 
elements $(\mathbf{\Delta}_\pm)_{(j-1)N+k}=\RDelta_{0j}\pm\RDelta_k$, 
$(\boldsymbol\sigma(u,\Rt))_{(j-1)N+k}=\bar{\sigma}_{j,k}(u,\Rt)$ and 
$(\mathbf{g})_{(j-1)N+k}=\RDelta_{0\delta}\RDelta_\delta
G_0(\RDelta_{0j})G(\RDelta_k)$ for $1\leq j\leq K$, $1\leq k\leq N$. With
these definitions \eref{sigma_stage2_laplace_discrete} becomes
\begin{equation}\label{sigma_stage2_laplace_matrix1}
\fl
\frac{\partial}{\partial \Rt}\boldsymbol\sigma(u,\Rt)
=-\rmi D(\mathbf{\Delta}_+) \boldsymbol\sigma(u,\Rt)
-\frac{1}{u}\mathbf{h}^{(KN)}\mathbf{g}^T \boldsymbol\sigma(u,\Rt)
+\frac{\rmi}{u}E_{\rm in}(\Rt+\Rtaup)\mathbf{h}^{(KN)}.
\end{equation}
Defining $M_2(u)=-\rmi D(\mathbf{\Delta}_+)
-\mathbf{h}^{(KN)}\mathbf{g}^T/u$ we can write
\eref{sigma_stage2_laplace_matrix1} as
\begin{equation}\label{sigma_stage2_laplace_matrix2}
\frac{\partial}{\partial \Rt}\boldsymbol\sigma(u,\Rt)
=M_2(u)\boldsymbol\sigma(u,\Rt)
+\frac{\rmi}{u}E_{\rm in}(\Rt+\Rtaup)\mathbf{h}^{(KN)}
\end{equation}
and the solution at the end of stage~2 is given by
\begin{equation}\label{sigma_in_the_end_of_stage2}
\fl
\boldsymbol\sigma^{(2)}(u)
=\frac{\rmi}{u}\int_0^{\Rtaud} \exp(M_2(u)\Rti)\mathbf{h}^{(KN)}
E_{\rm in}(\RtauR-\Rti)\rmd\Rti
+\exp(M_2(u)\Rtaud)\boldsymbol\sigma^{(1)}(u)
\end{equation}
with $\tauR=\taup+\taud$ as defined earlier.
Here we have used the initial condition $\boldsymbol\sigma^{(1)}(u)$ which in 
the vector notation has elements 
$(\boldsymbol\sigma^{(1)}(u))_{(j-1)N+k}=(\mathbf{P}(u,\Rtaup))_j$ for 
$1\leq j\leq K$ and $1\leq k\leq N$ with $\mathbf{P}(u,\Rt)$ given by 
\eref{P_stage1_laplace_matrix2_solution2}.

\subsection{Stage 3}
To describe stage 3 we could in principle have used the same equations as for 
stage 1 but then we would have lost the information about how the individual 
frequency components of the controlled broadening evolved. Hence doing this 
would not have allowed us to describe the rephasing in stage 4. Instead we use 
an equation similar to \eref{sigma_stage2_laplace_discrete} and set 
$\RDelta_k=0$. Here we do not have an incoming field so that the evolution is 
given by
\begin{equation}\label{sigma_stage3_laplace_discrete}
\fl
\frac{\partial}{\partial \Rt}\sigma_{jk}(u,\Rt)
=-\rmi\RDelta_{0j} \sigma_{jk}(u,\Rt)
-\frac{1}{u}\sum_{j'=1}^K \sum_{k'=1}^N
G_0(\RDelta_{0j'}) G(\RDelta_k')\RDelta_{0\delta}\RDelta_\delta
\sigma_{j'k'}(u,\Rt).
\end{equation}
We define a vector $\mathbf{\Delta}$ with elements 
$(\mathbf{\Delta})_{(j-1)N+k}=\RDelta_{0j}$ for $1\leq j\leq K$, 
$1\leq k\leq N$ and the corresponding matrix 
$M_3(u)=-\rmi D(\mathbf{\Delta})-\mathbf{h}^{(KN)}\mathbf{g}^T/u$. Then we can 
write \eref{sigma_stage3_laplace_discrete} as
\begin{equation}\label{sigma_stage3_laplace_matrix}
\frac{\partial}{\partial \Rt}\boldsymbol\sigma(u,\Rt)
=M_3(u)\boldsymbol\sigma(u,\Rt)
\end{equation}
so that the solution at the end of stage 3 is given by
\begin{equation}\label{sigma_in_the_end_of_stage3}
\boldsymbol\sigma^{(3)}(u)
=\exp(M_3(u)\Rtaud)\boldsymbol\sigma^{(2)}(u).
\end{equation}

\subsection{Stage 4}
For stage 4 we define 
$M_4(u)=-\rmi D(\mathbf{\Delta}_-)-\mathbf{h}^{(KN)}\mathbf{g}^T/u$ and the 
evolution of $\boldsymbol\sigma$ is then described by an equation of the same 
form as \eref{sigma_stage3_laplace_matrix} with the solution for $\sigma$ 
given by
\begin{equation}\label{sigma_in_stage4}
\boldsymbol\sigma(u,\Rt)=\exp(M_4(u)\Rt)\boldsymbol\sigma^{(3)}(u).
\end{equation}
From this expression we can find the electric field using the Laplace 
transformed and discretized version of \eref{E_slow_eqn_broadened} which can 
be written as
\begin{equation}\label{E_dicretized_laplace}
\bar{E}(u,\Rt)
=\frac{\rmi}{u}\mathbf{g}^T\boldsymbol\sigma(u,\Rt).
\end{equation}

We define the matrix
\begin{equation*}
J^{(KN\times KN)}(u)=\exp(M_3(u)\Rtaus)\exp(M_2(u)\Rtaud)
\end{equation*}
and the kernel
\begin{equation*}
\bar{k}_1(u,\Rt,\Rti')=-\frac{1}{u^2}\mathbf{g}^T
\exp(M_4(u)\Rt)\exp(M_3(u)\Rtaus)\exp(M_2(u)\Rti')
\mathbf{h}^{(KN)}.
\end{equation*}
Then we can use \eref{sigma_in_the_end_of_stage2}, 
\eref{sigma_in_the_end_of_stage3}, \eref{sigma_in_stage4} and
\eref{E_dicretized_laplace} to write the field as
\begin{equation}
\fl
\begin{array}{*3{>{\displaystyle}l}}
\bar{E}(u,\Rt)
&=&\int_0^{\Rtaud} \bar{k}_1(u,\Rt,\Rti') E_{\rm in}(\RtauR-\Rti')\rmd\Rti'\\
&&-\frac{1}{u^2}\mathbf{g}^T\exp(M_4(u)\Rt)J^{(KN\times KN)}(u)
\boldsymbol\sigma^{(1)}(u).
\end{array}
\end{equation}
Here the first term describes how the field incident in stage 2 is read out 
during stage 4 and the last term describes the readout of the field incident 
during stage 1.

Before we can write the second term in this expression as something involving 
the input field we need to express $\boldsymbol\sigma^{(1)}(u)$ in terms of 
$\mathbf{P}^{(1)}(u)$ so that we can use 
\eref{P_stage1_laplace_matrix2_solution2}. In the vector notation
$\mathbf{P}^{(1)}(u)$ is a $K$-dimensional vector while 
$\boldsymbol\sigma^{(1)}(u)$ is a $KN$-dimensional vector as defined in 
the end of \ref{stage2_appendix_subsection}. This definition means that every 
element in block number $j$ of $\boldsymbol\sigma^{(1)}(u)$ with length $N$ 
has the same value $(\mathbf{P}^{(1)}(u))_j$. Hence when the matrix 
$J^{(KN\times KN)}(u)$ is multiplied with $\boldsymbol\sigma^{(1)}(u)$ 
 each element of the resulting vector is given by
\begin{eqnarray*}
\fl
(J^{(KN\times KN)}(u)\boldsymbol\sigma^{(1)}(u))_j
=\sum_{k=1}^{KN}(J^{(KN\times KN)}(u))_{j,k}(\boldsymbol\sigma^{(1)}(u))_k\\
=\sum_{j'=1}^{K}\sum_{k'=1}^{N}(J^{(KN\times KN)}(u))_{j,(j'-1)N+k'}
(\boldsymbol\sigma^{(1)}(u))_{(j'-1)N+k'}.
\end{eqnarray*}
In the last line
$(\boldsymbol\sigma^{(1)}(u))_{(j'-1)N+k'}=(\mathbf{P}^{(1)}(u))_{j'}$ so that 
it can be taken out of the $k'$ summation. Hence if we define a matrix 
$J^{(KN\times K)}(u)$ with elements
\begin{equation*}
(J^{(KN\times K)}(u))_{j,k}
=\sum_{k'=1}^N (J^{(KN\times KN)}(u))_{j,(k-1)N+k'}
\end{equation*}
then
$J^{(KN\times KN)}(u)\boldsymbol\sigma^{(1)}(u)
=J^{(KN\times K)}(u)\boldsymbol P^{(1)}(u)$.
Then if we introduce another kernel
\begin{eqnarray*}
\bar{k}_2(u,\Rt,\Rti')=-\frac{1}{u^2}\mathbf{g}^T
\exp(M_4(u)\Rt)J^{(KN\times K)}(u)\exp(M_1(u)\Rti')
\mathbf{h}^{(K)}
\end{eqnarray*}
and take the inverse Laplace transform 
($u\rightarrow\Rz=1$) we can finally express the output field during stage 4 in 
terms of the input field
\begin{equation}\label{E_out_stage4}
\fl
E_{\rm out}(\Rt)
=\int_0^{\Rtaud} k_1(\Rt,\Rti')
E_{\rm in}(\RtauR-\Rti')\rmd\Rti'
+\int_0^{\Rtaup} k_2(\Rt,\Rti')
E_{\rm in}(\Rtaup-\Rti')\rmd t'
\end{equation}
where the kernels $k_1$ and $k_2$ are inverse Laplace transforms 
of $\bar{k}_1$ and $\bar{k}_2$ evaluated at $\Rz=1$.

\subsection{Stage 5}
In stage 5 the initial condition for $\sigma$ is given by 
\eref{sigma_in_stage4} with $\Rt=\Rtaud$. Defining
\begin{eqnarray*}
L^{(KN\times KN)}(u) = \exp(M_4(u)\Rtaud)\exp(M_3(u)\Rtaus),\\
B^{(KN\times KN)}(u) = \exp(M_4(u)\Rtaud)\exp(M_3(u)\Rtaus)\exp(M_2(u)\Rtaud),
\end{eqnarray*}
we can write the initial condition as
\begin{eqnarray*}
\fl
\boldsymbol\sigma^{(4)}(u)
=\frac{\rmi}{u}\int_0^{\Rtaud} L^{(KN\times KN)}(u)\exp(M_2(u)\Rti')
\mathbf{h}^{(KN)} E_{\rm in}(\RtauR-\Rti')\rmd\Rti'\\
+B^{(KN\times KN)}(u)\boldsymbol\sigma^{(1)}(u).
\end{eqnarray*}
The electric field can be found using the Laplace transformed and discretized 
version of \eref{E_slow_eqn_stage1} 
\begin{equation*}
\bar{E}(u,\Rt)
=\frac{\rmi}{u}\mathbf{g}_0^T\mathbf{P}(u,\Rt),
\end{equation*}
where $\mathbf{P}(u,\Rt)=\exp(M_1(u)\Rt)\mathbf{P}^{(4)}(u)$ is the solution 
to \eref{P_stage1_laplace_matrix2} with $E_{\rm in}=0$. 
The initial condition $\mathbf{P}^{(4)}(u)$ is defined by
\eref{P_defined_in_terms_of_sigma}. In vector notation this translates to 
blockwise summation of $\boldsymbol\sigma^{(4)}(u)$ with respect to the index 
of the controlled broadening weighted by the appropriate value of the 
controlled broadening distribution. We can apply this blockwise summation to 
the matrices $L^{(KN\times KN)}(u)$ and $B^{(KN\times KN)}(u)$ directly. Hence 
we define $L^{(K\times KN)}(u)$ and $B^{(K\times K)}(u)$ with
\begin{eqnarray*}
(L^{(K\times KN)}(u))_{j,k}=\sum_{j'=1}^N
G(\RDelta_{j'})\RDelta_\delta (L^{(KN\times KN)}(u))_{(j-1)N+j',k},\\
(B^{(K\times K)}(u))_{j,k}=\sum_{j'=1}^N \sum_{k'=1}^N
G(\RDelta_{j'})\RDelta_\delta (B^{(KN\times KN)}(u))_{(j-1)N+j',(k-1)N+k'}
\end{eqnarray*}
where we also apply the same reduction for the matrix $B^{(KN\times KN)}(u)$ 
as for $J^{(KN\times KN)}(u)$ previously. We can now use kernels
\begin{eqnarray*}
\bar{k}_3(u,\Rt,\Rti')=-\frac{1}{u^2}\mathbf{g}_0^T
\exp(M_1(u)\Rt)L^{(K\times KN)}(u)\exp(M_2(u)\Rti')
\mathbf{h}^{(KN)},\\
\bar{k}_4(u,\Rt,\Rti')=-\frac{1}{u^2}\mathbf{g}_0^T
\exp(M_1(u)\Rt)B^{(K\times K)}(u)\exp(M_1(u)\Rti')
\mathbf{h}^{(K)}.
\end{eqnarray*}
to express the output electric field in terms of the input field 
$E_{\rm in}$. Taking the inverse Laplace transform ($u\rightarrow\Rz=1$) the 
output field is
\begin{equation}\label{E_out_stage5}
\fl
E_{\rm out}(\Rt)
=\int_0^{\Rtaud} k_3(\Rt,\Rti')
E_{\rm in}(\RtauR-\Rti')\rmd\Rti'
+\int_0^{\Rtaup} k_4(\Rt,\Rti')
E_{\rm in}(\Rtaup-\Rti')\rmd\Rti'.
\end{equation}
where the kernels $k_3$ and $k_4$ are inverse Laplace transforms of $\bar{k}_3$ 
and $\bar{k}_4$ evaluated at $\Rz=1$.

\subsection{Concluding remarks}
With the above results we have found the sought relation between the input and 
output field. The final expressions in \eref{E_out_stage4} and 
\eref{E_out_stage5} give the output field during stages 4 and 5 respectively. 
They can be combined into a single expression \eref{E_out_combined} if we 
define
\begin{equation*}
K_E(\Rt,\Rti')=\left\{\begin{array}{ll}
k_1(\Rt,\Rti') &{\rm for}\quad \Rt\leq\Rtaud,\Rti'\leq\Rtaud;\\
k_2(\Rt,\Rti'-\Rtaud) &{\rm for}\quad \Rt\leq\Rtaud,\Rti'>\Rtaud;\\
k_3(\Rt-\Rtaud,\Rti') &{\rm for}\quad \Rt>\Rtaud,\Rti'\leq\Rtaud;\\
k_4(\Rt-\Rtaud,\Rti'-\Rtaud) &{\rm for}\quad \Rt>\Rtaud,\Rti'>\Rtaud.
\end{array}\right.
\end{equation*}
Here the 
shifts of the arguments of the kernels are performed because we go from 
considering $E_{\rm out}$ to be output fields only in stage~4 or only in 
stage~5 (where time was reset at the beginning of each stage) to having a 
single function $E_{\rm out}$ giving outgoing field in both stages. In actual 
simulations we use expressions \eref{E_out_stage4} and \eref{E_out_stage5} 
directly, but expression \eref{E_out_combined} is a more convenient 
formulation for the arguments made in the main text.

For inversion of the Laplace transform we use numerical integration along the 
Talbot contour \cite{Weideman06}. Since this method has exponential 
convergence, only a few discrete Laplace moments $u$ are needed. This makes it 
possible to compute and store certain intermediate results for each $u$. For 
example the kernels $k_1$, $k_2$, $k_3$ and $k_4$ can be computed much faster 
if the matrices $M_1(u)$, $M_2(u)$ and $M_4(u)$ are diagonalized. Hence we 
compute and store the eigenvalues and eigenvectors of those matrices, so that 
matrix exponentials can be evaluated by exponentiating the eigenvalues instead 
of the whole matrix.

\bibliographystyle{unsrt}
\bibliography{references}
\end{document}